\documentstyle[multicol,prb,aps,epsf]{revtex}
\newcounter{mycount}
\newcommand{\myroman}[1]{\setcounter{mycount}{#1}\roman{mycount}}
\begin{document}
\draft
\title{Detailed analysis of scanning tunneling microscopy images of the Si(001) reconstructed surface with buckled dimers }
\author{H. Okada, Y. Fujimoto, K. Endo, K. Hirose, and Y. Mori}
\address{Department of Precision Science and Technology, Osaka University, Suita, Osaka 565-0871, Japan}
\date{\today}
\maketitle
\begin{abstract}
The adequate interpretation of scanning tunneling microscopy (STM) images of the clean Si(001) surface is presented. We have performed both STM observations and {\it ab initio} simulations of STM images for buckled dimers on the clean Si(001) surface. By comparing experimental results with theoretical ones, it is revealed that STM images depend on the sample bias and the tip-sample separation. This enables us to elucidate the relationship between the corrugation in STM images and the atomic structure of buckled dimers. Moreover, to elucidate these changes, we analyze details of the spatial distributions of the $\pi$, $\pi^{\ast}$ surface states and $\sigma$, $\sigma^{\ast}$ Si-Si bond states in the local density of states which contribute to STM images.

\end{abstract}

\pacs{}

\begin{multicols}{2}
\narrowtext

\section{INTRODUCTION}
Scanning tunneling microscopy (STM) has provided much useful information on surface configurations, with atomic resolution. The clean Si(001) surface is vigorously studied by STM because of the great fundamental and technological importance of semiconductor devices. In the STM images of the 2$\times$1 structure on the terrace, silicon dimers appear to be symmetric at room temperature due to the time average of the flip-flop motion of buckled dimers, while at temperatures below $\sim$200 K, the stable configuration is the $c(4$$\times$$2)$ or $p(2$$\times$$2)$ structure consisting of asymmetric dimer rows.\cite{wolkow,shigekawa,yokoyama2} For the $c(4$$\times$$2)$ structure, the dependence of the STM image on the bias voltage at 80 K has been studied in detail.\cite{hata} On the other hand, the behavior of the dimers at step edges differs from that on the terrace. For example, the dimers at $S_A$ steps are observed to be buckled at both low and room temperatures.\cite{tromp} However, detailed observation of the $S_A$ steps has not been exhaustively carried out, since filled- and empty-state images at only a few bias voltages have been reported.\cite{tochihara}

It is known that STM images reflect electronic structures rather than atomic geometries. Consequently, bias-dependent STM images result from the variation of the spatial distribution of the local density of states (LDOS) which is a function of energy. Therefore, to interpret STM images properly, the spatial distributions of the LDOS must be calculated strictly by {\it ab initio} methods. Some authors have attempted to compare experimental STM images with theoretical ones for the Si(001) reconstructed surface.\cite{hata,kageshima} Kageshima and Tsukada carried out the simulation of polarity-dependent STM images of Si(001). However, their investigation was based on the semi-empirical method with linearly combined atomic orbitals.\cite{kageshima} Hata et al. pointed out that the $\pi$, $\pi^{\ast}$ surface states and the $\sigma$, $\sigma^{\ast}$ bulk ones contribute to the STM images, by calculating the spatial distributions of those states localized near the outermost atoms.\cite{hata} These results emphasize that a reproduction of the STM image from the calculated spatial distribution of the LDOS is very important. But such research is still limited at present.

In this paper, we present an improved interpretation of STM images and provide a more profound understanding of the electronic structure for the clean Si(001) surface. STM observations are performed on the buckled dimers at the $S_A$ step of the clean Si(001) surface. They are strongly dependent on the applied bias voltage and the tip-sample separation, which is confirmed by {\it ab initio} simulations. The STM images will be interpreted as isosurfaces of the spatial distribution of the LDOS in the vacuum region between a tip and a sample, which is greatly different from the behavior of the LDOS near the outermost atoms. Moreover, the calculated LDOS shows how the spatial distributions of the characteristic surface states ($\pi$, $\pi_{1}^{\ast}$, and $\pi_{2}^{\ast}$ dangling-bond states) and bulk states ($\sigma$ and $\sigma^{\ast}$ Si-Si bond states) contribute to the bias dependence of the STM images. 

\section{Methods of STM observation and theoretical calculation}
\subsection{STM observations}
Experiments were performed in an ultrahigh-vacuum (UHV) chamber equipped with STM. The base pressure was less than 1$\times$$10^{-8}$ Pa. The samples were Si(001) (B-doped, 0.2--0.5 $\Omega$cm) wafers. They were degassed at 900 K overnight, flashed to 1450 K for 30 sec and then cooled to room temperature to prepare the clean Si(001)-2$\times$1 surface. The tips were prepared by the electrochemical etching of a polycrystalline tungsten wire and cleaned by electron bombardment in UHV prior to use. The constant-current STM images were obtained at room temperature.

\subsection{Theoretical calculations}
The {\it ab initio} electronic-states calculations were performed by the self-consistent pseudopotential method within the density-functional theory.\cite{kohn} The exchange-correlation interaction was treated by the Ceperley-Alder form in the local-density approximation (LDA).\cite{perdew} The norm-conserving Troullier--Martins pseudopotentials\cite{tm} were used in a separable nonlocal form.\cite{separable} The wave functions were expanded in a plane-wave basis set with a  cutoff energy of 24 Ry.

The Si(001)-$p(2$$\times$$2)$ surface with asymmetric dimers was modeled as a repeated slab consisting of five silicon layers, the lowest of which was terminated by hydrogen atoms, and the width of the vacuum region was equivalent to that of 12 silicon layers. The four outermost silicon layers were fully optimized by the {\it ab initio} molecular-dynamics method.\cite{car}

Many theoretical STM images have been generated in the Tersoff-Hamann approximation.\cite{th,bardeen,chen} This method is actually valid for many systems despite its extreme simplicity.\cite{tunnel} In this scheme, the tunneling current is proportional to the surface LDOS at the tip position integrated over the energy range restricted by the applied bias voltage. We calculated the LDOS $\rho_s (x,y,z;\epsilon)$ at spatial points $(x,y,z)$ and energy $\epsilon$ by sampling the 100 $k$-points of the surface Brillouin zone (SBZ). The STM images were generated from the isosurface of the spatial distribution formed by the integration of the LDOS over the energy range from Fermi energy $E_F$ to $E_F$-$eV$ with applied voltage $V$, i.e., $\int_0^{eV} \rho_s (x,y,z;E_F-eV+\epsilon) d \epsilon$. Hereafter, we refer to this `energy-integrated LDOS' as the `EI-LDOS'. Thereby the computed STM images correspond to experimental constant-current images.

\section{Results and discussion}
\subsection{Optimized atomic structure and electronic structure of the Si(001)-$p(2$$\times$$2)$ surface}
The atomic structure of the Si(001)-$p(2$$\times$$2)$ surface was optimized by minimizing the total energy. The dimer bond length of 2.35 \AA, which is the same length as the bond length in bulk silicon, and the buckling angle of $19.0^{\circ}$ were found. These values are similar to the dimer bond length of 2.38 \AA\ and the buckling angle of $18.4^{\circ}$ for the $p(2$$\times$$2)$ reconstruction reported by Shkrebtii {\it et al}.\cite{shkrebtii} Our calculated dimer bond length is longer than the 2.28 \AA\ found by Ramstad {\it et al}.\cite{ramstad} and 2.31 \AA\ by Gunnella {\it et al}.\cite{gunnella} The backbond lengths of the upper and lower atoms are 2.38 and 2.33 \AA, respectively, which are longer than the of 2.34 and 2.31 \AA\ reported by Ramstad {\it et al}.\cite{ramstad} and 2.33 and 2.30 \AA\ by Gunnella {\it et al}.\cite{gunnella}

\begin{figure}
\epsfbox{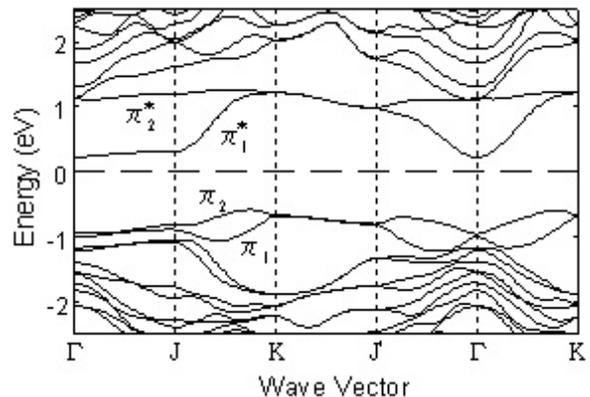}
\caption{Band structure of the Si(001)-$p(2$$\times$$2)$ surface optimized by molecular dynamics. Eigenvalues in the empty states are rigidly shifted by 0.65 eV. To match our STM results, the Fermi energy $E_F$ position is determined at 0.6 eV above the valence band top.}
\label{band-structure}
\end{figure}

We calculated the band structure for the optimized Si(001)-$p(2$$\times$$2)$ surface. This is formed at 100 equidistant points along the Bloch wave vector ${\bf k}_{\parallel}$ in two-dimensional (001) SBZ. Figure~\ref{band-structure} shows the band structure for the Si(001)-$p(2$$\times$$2)$ surface along high-symmetry directions of the SBZ. The single-particle calculation within the LDA is well known to underestimate the band gap. Although the band gap is calculated as 0.15 eV, all empty states of the band structure must shift rigidly upwards in energy by 0.65 eV.\cite{godby,northrup1,kipp} This value was determined from the results of angle-resolved photoemission,\cite{johansson1} angle-resolved inverse photoemission,\cite{johansson2} and scanning tunneling spectroscopy (STS).\cite{yokoyama} The position of the Fermi energy ($E_F$) is determined at about 0.6 eV above the valence-band top and about 0.2 eV below the conduction-band bottom in order to coincide with our STM results.

Filled $\pi$ and empty $\pi^{\ast}$ surface states separate into two filled ($\pi_{1}$, $\pi_{2}$) and empty ($\pi_{1}^{\ast}$, $\pi_{2}^{\ast}$) surface states in the bulk band gap, respectively. This feature is consistent with the results of other calculations.\cite{ramstad,zhu,northrup} The filled $\pi_{1}$ and $\pi_{2}$ states disperse from --1.2 to --0.6 eV. Although the empty $\pi_{1}^{\ast}$ state widely spreads from +0.2 to +1.1 eV, the empty $\pi_{2}^{\ast}$ state spreads at around +1.1 eV. The $\pi_{1}$ and $\pi_{2}$ states, as well as the $\pi_{1}^{\ast}$ and $\pi_{2}^{\ast}$ states, overlap each other from the $K$ to the $J^{\prime}$ point (in the direction perpendicular to the dimer rows). 

\subsection{Comparison of experimental STM images with simulated ones}
\begin{figure}
\epsfbox{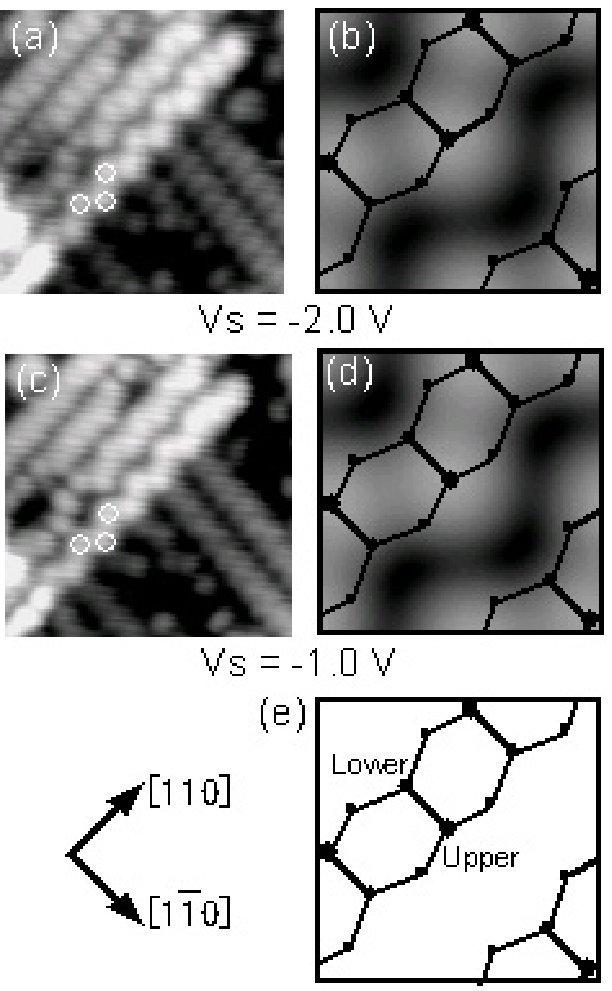}
\caption{Filled-state STM images (5$\times$5 nm$^{2}$) of the $S_A$ step on the Si(001)-2$\times$1 surface at room temperature (left column) and simulated ones of the Si(001)-$p(2$$\times$$2)$ surface (right column). These experimental images were obtained at sample biases of (a) --2.0 and (c) --1.0 V and at a tunneling current of 0.5 nA. These simulated images were generated from the isosurfaces of EI-LDOS [(b) 0 to --2.0 and (d) 0 to --1.0 eV] of 8.2$\times$10$^{-5}$ electrons/\AA$^3$. The maximal heights of these isosurfaces are (b) 4.9 and (d) 4.7 \AA\ from the upper atom. The atomic positions in these images are represented in (e) and the smallest filled circles are the second-layer atoms.}
\label{f-images}
\end{figure}

\begin{figure}
\epsfbox{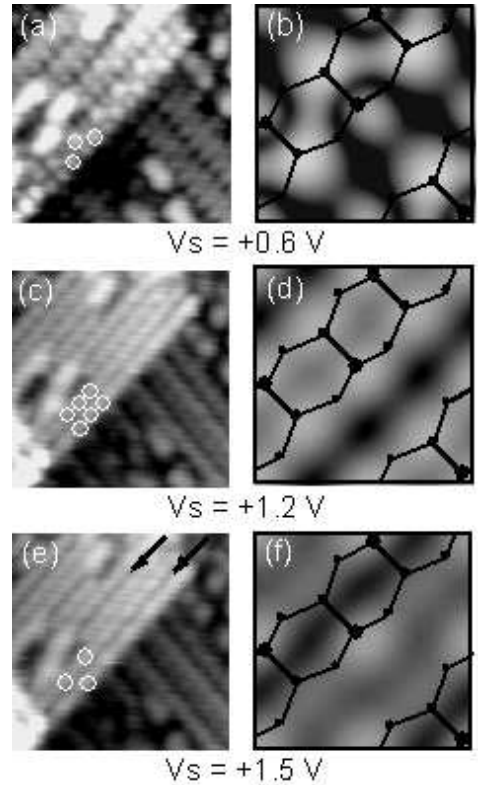}
\caption{Empty-state STM images (5$\times$5 nm$^{2}$) of the $S_A$ step on the Si(001)-2$\times$1 surface at room temperature (left column) and simulated ones of the Si(001)-$p(2$$\times$$2)$ surface (right column). These experimental images are obtained at sample biases of (a) +0.6, (c) +1.2, and (e) +1.5 V and at a tunneling current of 0.5 nA. These simulated images are generated at the isosurfaces of EI-LDOS [(b) 0 to +0.6, (d) 0 to +1.2, and (f) 0 to +1.5 eV] of 8.2$\times$10$^{-5}$ electrons/\AA$^3$. The maximal heights of these isosurfaces are (b) 3.4, (d) 4.2, and (f) 4.6 \AA\ from the upper atom. The atomic positions in these images are represented in Fig.~\ref{f-images}(e).}
\label{e-images}
\end{figure}

We aimed at obtaining the STM images of a buckled-dimer row at the $S_A$ step on the clean Si(001)-2$\times$1 surface under different sample-bias conditions. Figures~\ref{f-images} (filled states) and~\ref{e-images} (empty states) show the experimental STM images of the $S_A$ step and the simulated ones of the Si(001)-$p(2\times2)$. All STM images in Figs.~\ref{f-images} and~\ref{e-images} were obtained at a tunneling current of 0.5 nA, and simulated images were generated by EI-LDOS of 8.2$\times10^{-5}$ electrons/\AA$^3$. The simulated images accurately reproduce the experimental results.

In the experimental images of filled states [Figs.~\ref{f-images}(a) and~\ref{f-images}(c)], the dimer row at the $S_A$ step makes a zigzag pattern. In other words, one atom in a buckled dimer is brightly observed and the other is absent.\cite{wolkow,shigekawa,yokoyama2,hata,tromp,tochihara} The filled-state images do not change greatly in the range of bias voltage from --1.0 to --2.5 V, except for the difference in the corrugation height. The simulated filled-state images [Figs.~\ref{f-images}(b) and~\ref{f-images}(d)] imply that the bright protrusion of the buckled dimer in the experimental images corresponds to the upper atom of the buckled dimer.

On the other hand, the empty-state images of the buckled dimer at the $S_A$ step show dramatic changes at room temperature as the sample bias is increased. These images can be divided into three typical types depending on applied voltage: (\myroman{1}) below +1.1 V, (\myroman{2}) at about +1.2 V, and (\myroman{3}) from +1.3 to +2.0 V. The empty-state images in Fig.~\ref{e-images} were obtained for the same area, as shown in Fig.~\ref{f-images}. At a sample bias of +0.6 V in Fig.~\ref{e-images}(a), the zigzag corrugation reverses to that in the filled-state images in Figs.~\ref{f-images}(a) and \ref{f-images}(c).\cite{tochihara} This reversal is also observed on the terrace for the Si(001)-$c(4$$\times$$2)$ surface at low temperatures.\cite{yokoyama2,hata} Inside one buckled dimer in Fig.~\ref{e-images}, two protrusions are clearly distinguished. One is brighter than the other, which cannot be seen in the filled states [Figs.~\ref{f-images}(a) and~\ref{f-images}(c)].
The simulated STM image at a sample bias of +0.6 eV in Fig.~\ref{e-images}(b) coincides with the experimental one [Fig.~\ref{e-images}(a)]. The brighter protrusion in Fig.~\ref{e-images}(a) corresponds to the lower atom of the buckled dimer. 
When +1.2 V is applied to the sample, the experimental and simulated images show that the zigzag dimer row changes into a symmetric one [Figs.~\ref{e-images}(c) and~\ref{e-images}(d)]. At a sample bias of +1.5 V in Fig.~\ref{e-images}(e), the brighter oval-shaped protrusion is located at the opposite side to that at a sample bias of +0.6 V in the buckled dimer. 
The atomic protrusions align parallel to the dimer rows. In addition, a deep trough appears in the middle of the dimer row, as indicated by black arrows in Fig.~\ref{e-images}(e). 
This has been observed in the empty-state images at high biases at both room and low temperatures.\cite{hata,hamers} These features are accurately represented in the simulated image, as shown in Fig.~\ref{e-images}(f). The bright protrusion corresponds to the upper atom of the buckled dimer.

\begin{figure}
\epsfbox{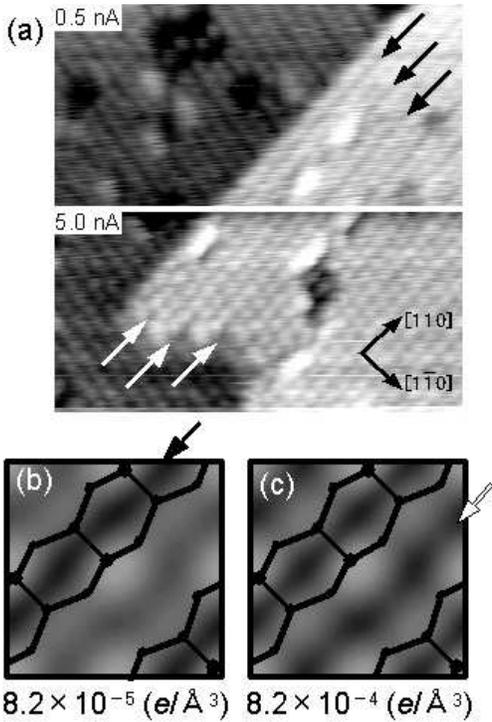}
\caption{Current-dependent STM images (9$\times$9 nm$^{2}$) of the Si(001)-2$\times$1 surface of the empty states at room temperature and the simulated image. (a) The experimental image is obtained at a sample bias of +1.5 V. In the upper and lower parts of the STM image, the tunneling current is 0.5 and 5.0 nA, respectively. (b) This image is equivalent to Fig.~\ref{e-images}(f). (c) The simulated image is generated by the isosurface of 8.2$\times$10$^{-4}$ electrons/\AA$^3$, which is one order larger than that adopted in Fig.~\ref{e-images}(f). The maximal height of this isosurface is 3.7 \AA\ from the upper atom.}
\label{current-stm}
\end{figure}

We also investigate the current dependence of buckled dimers and dimer rows on the Si(001) surface in STM images. Figure~\ref{current-stm}(a) shows STM images of the Si(001)-2$\times$1 surface at a sample bias of 1.5 V. In the middle of the image, the tunneling current was switched from 0.5 to 5.0 nA. In the upper (lower) area of the image, the tunneling current was 0.5 (5.0) nA. In the upper half of the figure, the deep trough appears in the center of the dimer row, as mentioned above. The deep trough exists in the center of the dimer row indicated by black arrows in Fig.~\ref{current-stm}(a) (at a tunneling current of 0.5 nA), as mentioned above. When the tunneling current is changed to 5.0 nA, the other trough appears between the dimer rows, as indicated by white arrows in Fig.~\ref{current-stm}(a). Furthermore, the two atoms in a dimer are clearly visible. Since the tip approaches the sample surface on switching the tunneling current from 0.5 to 5.0 nA, the LDOS at the nearer sample surface contributes to the tunneling current. In Fig.~\ref{current-stm}(c) the simulated image is generated as the isosurface of 8.2$\times$10$^{-4}$ electrons/\AA$^3$, which is one order larger than that adopted in Fig.~\ref{e-images}(f). This simulated image is consistent with the experimental one and indicates that STM images depend on the magnitude of the surface LDOS, i.e., the tunneling conductance.

Consequently, the simulation of STM images can support the interpretation of the experimental STM images, as shown in Figs.~\ref{f-images},~\ref{e-images}, and~\ref{current-stm}. The simulated ones generated by the EI-LDOS of 8.2$\times$10$^{-5}$ electrons/\AA$^{3}$ corresponded well with the experimental STM images obtained at the tunneling current of 0.5 nA. We can estimate the distance from the surface at which the spatial distribution of the EI-LDOS contributes to the tunneling current, i.e., the STM image. When the sample bias and the tunneling current are set to +1.5 and 0.5 (5.0) nA, the distance of the EI-LDOS from the surface contributing to the STM images is predicted to be 4.6 (3.7) \AA. Similarly, when the tunneling current is fixed to be 0.5 nA, the EI-LDOS at about 4.9, 4.7, 3.4, and 4.2 \AA\ from the surface contributes to the STM images at the sample biases of --2.0, --1.0, +0.6, and +1.2 V, respectively. As the absolute value of the bias voltage is increased or the tunneling current is decreased, the distance from the surface at which the surface LDOS contributes to the tunneling current becomes large. When tip-sample separations are assumed to be about twice as long as the distances estimated above, they agree with the experimental values \cite{tunnel}. To estimate the tip-sample separation exactly, both the tip and the sample must be taken into account in the calculation.

\subsection{LDOS and spatial distribution of each state in LDOS of the Si(001)-$p(2$$\times$$2)$ surface}
By calculating the LDOS at different sites on the Si(001)-$p(2$$\times$$2)$ surface, we demonstrate how the spatial distributions of the characteristic surface and bulk states in the LDOS contribute to the STM images.

\begin{figure}
\epsfbox{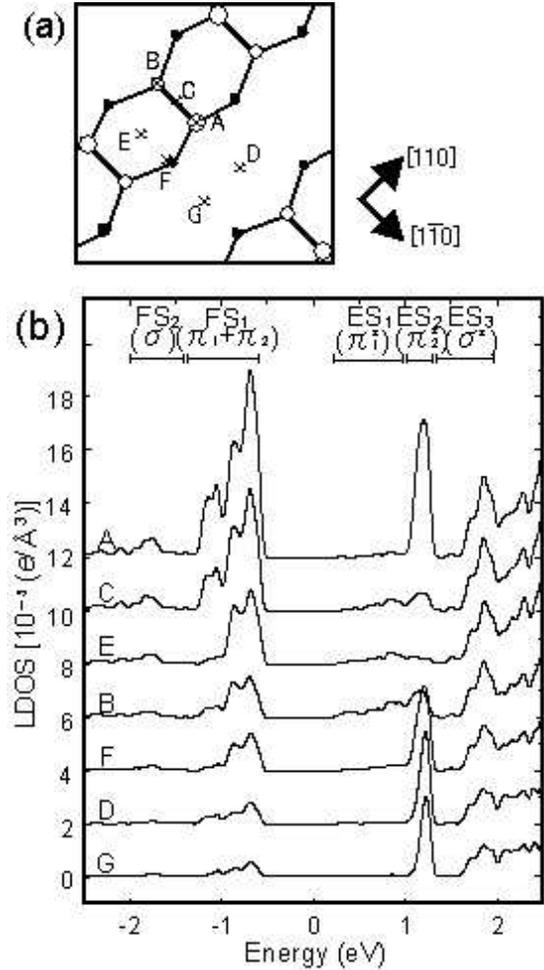}
\caption{Calculated LDOS of the Si(001)-$p(2$$\times$$2)$ surface at different sites. (a) Top view indicates the seven sites at which the LDOS are calculated. A, upper atom; B, lower atom; C, bridge; D, cave; E, pedestal; F, side bridge; and G, valley bridge. (b) The LDOS at these seven sites are generated at the same height of 4.3 \AA\ from the upper atom. The five characteristic peaks at --0.7, --1.7, +0.8, +1.2, and +1.8 eV are marked by FS$_1$, FS$_2$, ES$_1$, ES$_2$, and ES$_3$, respectively.}
\label{ldos}
\end{figure}

Figure~\ref{ldos} shows the LDOS at different sites on the Si(001)-$p(2$$\times$$2)$ surface. The LDOS in Fig.~\ref{ldos}(b) are generated at seven specific sites (sites A--G) in Fig.~\ref{ldos}(a) which are located at the height of 4.3 \AA\ above the upper dimer atom. In the filled states, we find a sharp peak at --0.7 eV and a broad one at --1.7 eV marked by FS$_1$ and FS$_2$ respectively, in Fig.~\ref{ldos}(b). These peaks appear similarly in every spectrum in Fig.~\ref{ldos}(b). By comparing these peaks with the band structure (Fig.~\ref{band-structure}), FS$_1$ and FS$_2$ mainly correspond to occupied [$\pi$ ($\pi_1$ and $\pi_2$)] dangling-bond states and the Si-Si bond ($\sigma$) state, respectively.  FS$_1$ and FS$_2$ at the upper atom [site A in Fig.~\ref{ldos}(a)] are the strongest, while they are the weakest at the cave and valley bridge sites [sites D and G in Fig.~\ref{ldos}(a)].

For the empty states, the LDOS have a small peak at around +0.8 eV (ES$_1$), a sharp one at +1.2 eV (ES$_2$), and a broad one at +1.8 eV (ES$_3$), as shown in Fig.~\ref{ldos}(b). From the band structure, ES$_1$ and ES$_2$ mainly correspond to unoccupied $\pi_{1}^{\ast}$ and $\pi_{2}^{\ast}$ dangling-bond states, respectively. The heights and the fine structures of ES$_1$ and ES$_2$ at each site greatly differ from one another, which causes a dramatic change in the empty-state STM images. ES$_1$ at the lower atom, bridge, and pedestal sites [sites B, C, and E in Fig.~\ref{ldos}(a)] are stronger than at other sites. ES$_2$ at the upper atom, cave, side bridge, and valley bridge sites (sites A, D, F, and G) is much sharper than at other sites. In contrast, ES$_2$ at the pedestal site (site E) decreases markedly. As described below, the analysis of the spatial distribution of each state is useful for understanding STM images.

We compare our calculated LDOS with previously reported STS measurements for the Si(001)-2$\times$1 surface.\cite{yokoyama,hamers,boland} The experimental tunneling spectra $(dI/dV)/(I/V)$ have some characteristic features. (\myroman{1}) The band gap is $\sim$0.6 eV. (\myroman{2}) In the filled states, a large peak is found around --0.7 eV. (\myroman{3}) In the empty states, two peaks are observed at around +0.6 and +1.3 eV. These features agree with those of our theoretical LDOS. We attribute the peaks at --0.7, +0.6, and +1.3 eV to $\pi$, $\pi_{1}^{\ast}$, and $\pi_{2}^{\ast}$, respectively. 

\begin{figure}
\epsfbox{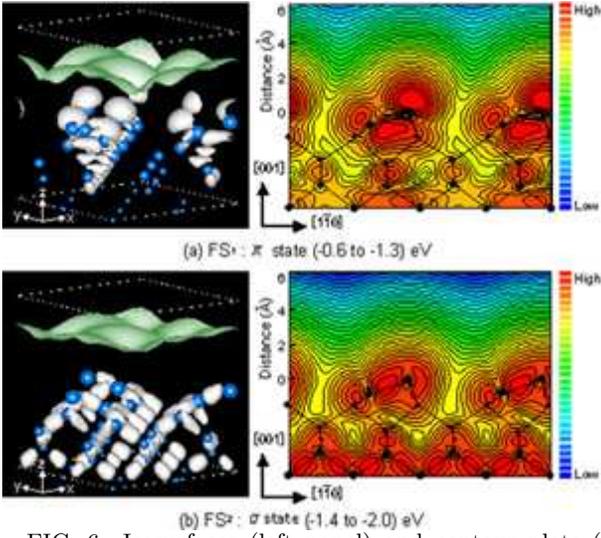}
\caption{Isosurfaces (left panel) and contour plots (right panel) of charge densities for the two characteristic peaks in the filled states: (a) --0.7 (--0.6 to --1.3) and (b) --1.7 (--1.4 to --2.0) eV. Isosurfaces (left panel) in the bulk (white) and the vacuum (at $\sim$4.3 \AA\ from the upper atom; green) are shown in the same cell; the height of silicon layers is represented by the size of the ball. Logarithmic contour maps (right panel) are along the cross section in the (110) plane along the $[1\bar{1}0]$ direction including the buckled dimer. The magnitude of the energy-integrated LDOS increases/decreases one order per four contour lines.}
\label{f-contour}
\end{figure}

We calculate the spatial distribution of each characteristic state by integrating the LDOS over the energy range in which peaks relating to the state spread are found. We also define this `energy-integrated LDOS' as `EI-LDOS'.
Figure~\ref{f-contour} shows the spatial distributions of EI-LDOS of the two characteristic peaks [(a) FS$_1$ and (b) FS$_2$] in the filled states. Figure.~\ref{f-contour}(a) shows that the charge transfer occurs from the lower atom to the upper one and the EI-LDOS is largely localized at the upper atom \cite{chadi1}. The maximum EI-LDOS in the vacuum region is located just above the upper atom. For FS$_2$ [Fig.~\ref{f-contour}(b)], the spatial distribution of the EI-LDOS is largely localized between Si-Si atoms and forms $\sigma$ bonds, which is greatly different from the $\pi$ surface state (FS$_1$). In contrast, in the vacuum region, the maximum EI-LDOS of FS$_2$ is again located above the upper atom, although the corrugation of contour lines for FS$_2$ is smaller than that for FS$_1$. This feature leads to the similarity of STM images at any sample bias in the filled state, as shown in Fig.~\ref{f-images}.

\begin{figure}
\epsfbox{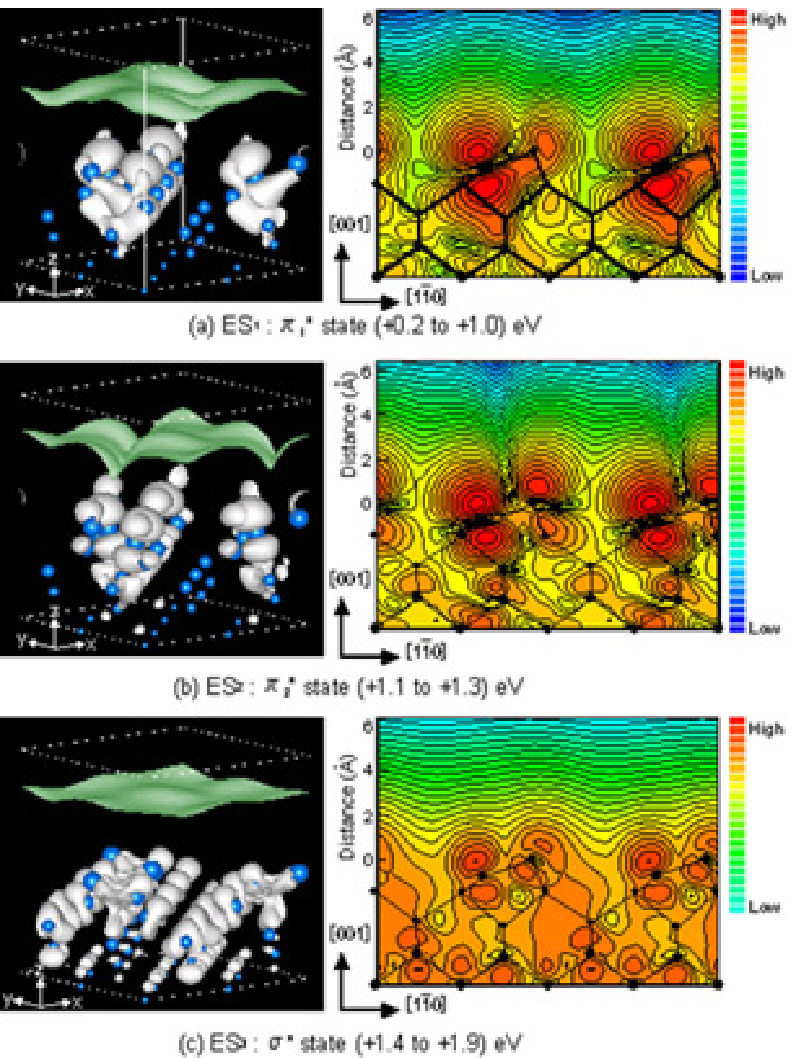}
\caption{Isosurfaces (left panel) and contour plots (right panel) of charge densities for the three characteristic peaks in the empty states: (a) +0.8 (+0.2--+1.0), (b) +1.2 (+1.1--+1.3), and (c) +1.8 (+1.4--+2.0) eV. The meanings of the symbols are the same as in Fig.~\ref{f-contour}.}
\label{e-contour}
\end{figure}

Figure~\ref{e-contour} shows the spatial distributions of EI-LDOS of the three peaks [(a) ES$_1$, (b) ES$_2$, and (c) ES$_3$] in the empty states. In Fig.~\ref{e-contour}(a), the EI-LDOS of ES$_1$ is localized at the lower atom of the buckled dimer. In the vacuum region, the corrugation of the EI-LDOS has two humps above a buckled dimer and a minimum at the center of the dimer. The hump is considerably larger above the lower atom than above the upper one. The $\pi_{1}^{\ast}$ contributes to the STM images at low biases (below +1.1 eV). 
On the other hand, the spatial distribution of ES$_2$ is extremely different from that of ES$_1$. As shown in Fig.~\ref{e-contour}(b), the deep minimum appears in the middle of a buckled dimer [sites C and E in Fig.~\ref{ldos}(a)] and the maximum of the EI-LDOS of ES$_2$ extends into the vacuum above the upper atom more than above the lower atom. These striking features of the $\pi_{2}^{\ast}$ state lead to the deep trough in the middle of a dimer row and the bright oval-shaped protrusion at the upper dimer atom in the STM images at the sample bias from +1.2 to +2.0 V, as shown in Figs.~\ref{e-images}(e) and \ref{e-images}(f). In addition, we also find that the EI-LDOS of ES$_2$ decays slower in vacuum at the cave, valley bridge, and side bridge sites (sites D, F, and G) than at the other sites. 
This feature explains the changes in the STM image in Fig.~\ref{current-stm}(a) when a tunneling current is switched from 0.5 to 5.0 nA at a sample bias of +1.5 V. In other words, since the tip-sample separation is shorter and the STM image reflects the spatial distribution of the LDOS nearer to the surface, the other trough emerges between dimer rows and each atom appears to be clearly resolved. 
ES$_3$ corresponds to the $\sigma^{\ast}$ antibonding bulk state. As shown in Fig.~\ref{e-contour}(c), in the bulk, this state distributes not only at dimer rows but also between dimer rows and generally does not localize near the dimer atoms, in contrast with the $\pi_{1}^{\ast}$ and $\pi_{2}^{\ast}$ states. In the vacuum region, the corrugation of the EI-LDOS extends slightly above the upper atom. When the sample bias is higher, the position of the spatial distribution forming the STM image becomes farther from the surface. The LDOS of the $\sigma^{\ast}$ state decays slower in vacuum than that of the $\pi_{1}^{\ast}$ and $\pi_{2}^{\ast}$ states. Therefore, we predict that the $\sigma^{\ast}$ state is evident in the STM images at biases higher than +2.1 V.

We have presented the features of the spatial distribution of each state in the LDOS and revealed how the surface and bulk states contribute to STM images.
We also demonstrated that the STM image of the clean surface can be interpreted as the isosurface of the spatial distribution of the EI-LDOS in the vacuum region, which is greatly different from that near the outermost atoms. 

\section{CONCLUSION}
We have performed STM observations and theoretical simulations of STM images for the buckled dimer on the clean Si(001) surface. By comparing the experimental results with theoretical ones, the relationship between STM images and the atomic structure of buckled dimers of the Si(001) surface was clarified. The spatial distributions of the surface and Si-Si bond states in the LDOS were investigated. For the filled states, the $\pi$ ($\pi_1$ and $\pi_2$) surface states contribute to the low-bias STM images and, as the sample bias is increased, the Si-Si bond ($\sigma$) state is added to the $\pi$ states. Since the corrugations of the spatial distributions of the $\pi$ and $\sigma$ states are similar in vacuum, the filled-state images are nearly identical to each other. On the other hand, the empty-state images change considerably with relative to the sample bias. The STM images obtained at low biases reflect the spatial distribution of the $\pi_1^{\ast}$ surface states. As the sample bias is increased, the $\pi_2^{\ast}$ surface state begins to dominate the STM images, and a deep trough appears at the center of the dimer at high biases above +1.3 eV. Moreover, when the tunneling current is set to be larger, the LDOS nearer to the sample surface contributes to the STM image. These analyses enable us to understand not only STM images but also the electronic structure of the clean Si(001) surface.

\acknowledgments
The authors thank H. Goto of Kyoto Institute of Technology for discussing simulation programs. They also thank K. Sugiyama, K. Inagaki, K. Arima, and T. Ishikawa for their great support and useful discussions. This work was supported by a Grant-in-Aid for COE Research (No.08CE2004) from the Ministry of Education, Science, Sports and Culture.

\end{multicols}

\begin{references}
\bibitem{tunnel}
The conventional STM images are measured at the tip-sample separations of 5--10 \AA.\cite{feenstra,ohnishi} By using the Tersoff-Hamann approximation, the simulated STM images have been generated at the separations of 2--4 \AA\ from outermost atoms, i.e., tip-sample separations\cite{engels,khare,kim,bass,lu,mchen} which are shorter than the experimental ones. When the behavior of the LDOS of the tip is similar to that of the sample, it is likely that the tunneling current depends more on the sample and the LDOS of the sample and the tip around the ``middle of the tip-sample distance'', at which the magnitudes of these LDOS dominate the tunneling conductance, rather than that at the ``tip'' position. The tip-sample separation is estimated to be about twice as long as that adopted in the Tersoff-Hamann theory and is close to the typical experimental separation.
\end{references}
\end{document}